\documentclass[sigconf]{acmart}
\usepackage{changepage}
\usepackage{tabularx}
\usepackage{hyperref}
\usepackage{multicol}
\usepackage{multirow}
\usepackage[symbol]{footmisc}

\usepackage{tablefootnote}
\usepackage{threeparttable}

\usepackage[table]{xcolor} % Enable color in tables
\usepackage[most]{tcolorbox}
\usepackage{colortbl}      % For coloring table rows/cells

\usepackage{xcolor}
\definecolor{darkorange}{RGB}{153,76,0}
\definecolor{darkblue}{RGB}{0,101,203}
\definecolor{tealcolor}{RGB}{0,153,153}
\newcommand{\speech}[1]{{\textcolor{darkorange}{\textit{“#1”}}}}
\newcommand{\behavior}[1]{{\textcolor{darkblue}{[#1]}}}
\newcommand{\keyboard}[1]{{\textcolor{tealcolor}{\textbf{\texttt{#1}}}}}
\AtBeginDocument{%
  }

\setcopyright{acmlicensed}
\copyrightyear{2026}
\acmYear{2026}
\acmDOI{XXXXXXX.XXXXXXX}
\acmConference[SIGCSE TS 2026]{Proceedings of the 57th ACM
Technical Symposium on Computer Science Education V. 1 (SIGCSE TS 2026)}{February 18--21, 2026}{St. Louis, MO}
\acmISBN{978-1-4503-XXXX-X/XXXX/XX}

\begin{document}

\title{Exploring Student Choice and the Use of Multimodal Generative AI in Programming Learning}


\author{Xinying Hou}
\authornote{Both authors contributed equally to this research.}
\affiliation{%
  \institution{University of Michigan}
  \city{Ann Arbor}
  \country{United States}}
\email{xyhou@umich.edu}

\author{Ruiwei Xiao}
\affiliation{%
  \institution{Carnegie Mellon University}
  \city{Pittsburgh}
  \country{United States}}
\email{ruiweix@andrew.cmu.edu}
\authornotemark[1]

\author{Runlong Ye}
\affiliation{%
  \institution{University of Toronto}
  \city{Toronto}
  \country{Canada}}
\email{harryye@cs.toronto.edu}

\author{Michael Liut}
\affiliation{%
  \institution{University of Toronto Mississauga}
  \city{Mississauga}
  \country{Canada}}
\email{michael.liut@utoronto.ca}

\author{John Stamper}
\affiliation{%
  \institution{Carnegie Mellon University}
  \city{Pittsburgh}
  \country{United States}}
\email{jstamper@cmu.edu}

\renewcommand{\shortauthors}{Xinying Hou et al.}

\begin{abstract}
  The broad adoption of Generative AI (GenAI) is impacting Computer Science education, and recent studies found its benefits and potential concerns when students use it for programming learning. However, most existing explorations focus on GenAI tools that primarily support text-to-text interaction. With recent developments, GenAI applications have begun supporting multiple modes of communication, known as multimodality. In this work, we explored how undergraduate programming novices choose and work with multimodal GenAI tools, and their criteria for choices. We selected a commercially available multimodal GenAI platform for interaction, as it supports multiple input and output modalities, including text, audio, image upload, and real-time screen-sharing. Through 16 think-aloud sessions that combined participant observation with follow-up semi-structured interviews, we investigated student modality choices for GenAI tools when completing programming problems and the underlying criteria for modality selections. With multimodal communication emerging as the future of AI in education, this work aims to spark continued exploration on understanding student interaction with multimodal GenAI in the context of CS education.
\end{abstract}




\maketitle

\section{Introduction}
Computing education is increasingly impacted by the development of Generative AI (GenAI) technologies. Previous studies have explored how students interact with GenAI applications in the context of programming education \cite{Adams2024From,xiao2024exploring}. However, as GenAI technologies continue to develop, while the mainstream remains focused on text-to-text interaction, support for multimodal interactions is becoming increasingly common. Previous studies have evaluated the technical ability of multimodal GenAI models in solving visually heavy programming problems and tasks with text-plus-image or image-only input \cite{hou2024more, gutierrez2025seeing}. Latest studies also tested the capabilities in solving multimodal CS exercises \cite{ouh2025evaluating}. While the evaluation of multimodal GenAI models' capability is extensively tested, students’ modality choices in student-GenAI interactions and their underlying reasons are still underexplored. 

Although GenAI applications incorporating multimodal features are still emerging in educational research, a substantial body of work on multimodal learning already exists. In educational contexts, multimodal learning refers to instructional environments that engage learners through multiple sensory modalities (e.g., visual, auditory) \cite{sankey2010engaging}. This concept builds on Mayer’s Cognitive Theory of Multimedia Learning \cite{mayer2005cognitive}, which emphasizes the complementary and integrated use of multiple modalities to enhance understanding. Prior research has explored multimodal learning across diverse educational contexts (e.g., programming education \cite{grafsgaard2013modeling}, collaborative learning \cite{yan2024evidence}), demonstrating its effectiveness in supporting learner engagement and instructional interventions \cite{grafsgaard2013modeling, buffardi2022integrating, di2023rise} through enriched forms of instructional delivery.

% To leverage the benefits of multimodal learning, recent advances in generative AI, and the technical readiness for integrating the two, 
In this work, we applied Google AI Studio\footnote{https://aistudio.google.com/} as the GenAI interaction platform, which features two primary components: Chat and Stream Realtime. Chat allows multiple input modalities, such as text and image/file, but its main output modality is text only (including formatted code). Stream Realtime allows various input and output modalities. Its input modalities include screen sharing, audio, text, and images, and the output modalities are audio and the corresponding transcript. We conducted a think-aloud study followed by semi-structured interviews with 16 undergraduate programming novices. Students were allowed to choose their preferred modality to solve two types of short programming tasks. Throughout this study, we aimed to explore two main RQs:

\begin{adjustwidth}{2.5em}{}
\begin{enumerate}
    \item [\textbf{RQ1}:] What input-output modalities do students choose when solving short programming tasks, and how do they use it?
    \item [\textbf{RQ2}:] What do students consider when choosing modalities to interact with AI for programming learning?
\end{enumerate}
\end{adjustwidth}
\section{Related Work}
\paragraph{AI-based Educational Tools in CS: a Text-Focused Approach.} The rapid advancement of large language models (LLMs) has initiated a surge of research in computing education \cite{raihan2025large,cambaz2024use}. The community responses are mixed, recognizing the disruptive potential \cite{becker2023programming} while raising concerns about academic integrity, overreliance, and incorrect output \cite{kazemitabaar2023novices,kazemitabaar2024codeaid}. The shift toward pedagogical use of LLMs led to custom educational tools with ``pedagogical guardrails'', such as offering scaffolded assistance without revealing direct solutions \cite{kazemitabaar2024codeaid,hou2024codetailor} and improved student guidance strategies \cite{kumar2024supporting}. More ways to teach prompt crafting have also appeared, developing students' prompting skills via new exercises \cite{denny2024prompt} and through dedicated interactive instruction \cite{xiao2025improvingstudentaiinteractionpedagogical}. Besides these, studies on student AI help-seeking have provided crucial insights, alongside broader investigations of student perceptions on conversational chatbots used for programming learning \cite{Adams2024From, kazemitabaar2023novices}. The field is thus moving beyond technical feasibility to a nuanced exploration of how LLMs can serve CS education better. The research to date, however, has been predominantly focused on text-, prompt-, and code-based interactions. While this work provides a critical foundation, the recent emergence of capable multimodal models, which can interpret visual and auditory inputs, is beginning to enable novel programming interactions.

\vspace{0.5mm}

\paragraph{Emergence of Coding with Multimodal GenAI Models.} An emerging research area is on frontier multimodal models (e.g., GPT-o3, Gemini 2.5, Claude 4) that translate visual and speaking inputs into functional code. Research is rapidly demonstrating its technical feasibility across several modalities. These advanced models can process a diverse range of inputs, such as images, sketches, voice, live videos, and can respond to various formats, including code, natural language explanations, or spoken responses\footnote{\href{https://techcrunch.com/2024/09/24/openai-rolls-out-advanced-voice-mode-with-more-voices-and-a-new-look/}{TechCrunch: OpenAI rolls out Advanced Voice Mode with more voices and a new look}}. Within the programming domain, early exploration has focused on visual-to-code generation. For instance, models can now interpret static images of diagrams to generate code for formats like UML \cite{Bates_2025} or solve visual educational exercises like Parsons puzzles and data structure tasks \cite{hou2024more, gutierrez2025seeing}. This capability extends to translating UI sketches into runnable webpages, a task that is being systematically improved via open benchmarks \cite{jain2019sketch2codetransformationsketchesui, li2024sketch2code}. The interaction can be more dynamic, with tools that interpret freeform sketches drawn directly onto the code editor to perform iterative modifications \cite{yen2025code}. Auditory inputs are also being explored, primarily for using voice commands to create or manipulate code, which offers promise for accessibility \cite{wagner2012programming}, and has prompted research into programmers' natural speaking patterns ~\cite{nowrin2023programming}.

However, existing research has focused almost solely on technical performance, largely overlooking the pedagogical implications and the student experience. Our work begins to address this critical gap by shifting the focus from model capability to a student-centric exploration of how these unique modalities are perceived and used.

\vspace{0.5mm}

\paragraph{Theoretical Foundations for Multimodal Learning.} Our investigation is grounded in foundational learning theories that multimodal approaches can enhance computing education. Given programming's high mental demand, Cognitive Load Theory (CLT) suggests minimizing extraneous load; multimodal tools achieve this by allowing students to use voice and screenshare instead of manually describing complex code states, freeing up working memory \cite{SWELLER1988257, clt-review}. This cognitive offloading is complemented by Mayer’s Cognitive Theory of Multimedia Learning (CTML), which reasons that learning is most effective when both visual (seeing code) and auditory (hearing the AI) channels are engaged simultaneously \cite{mayer2005cognitive}. Finally, the theory of situated cognition argues for learning in authentic contexts \cite{brown1989situated}, which a multimodal AI enables by operating directly in a student's programming environment as a ``More Knowledgeable Other'' providing precisely scaffolded support \cite{vygotsky1978mind}.
\vspace{-4mm}

\section{Methods}
To answer RQs, we conducted a study with undergraduate students recruited from a large public university in North America. We distributed a recruitment message to students who had just completed an introductory-level undergraduate Python course (CS1) in Spring 2025. The observations were conducted remotely through Zoom, with each session lasting approximately 90 minutes. Each student received a \$20/hour gift card after completing the study. The study is approved by the Institutional Review Board (IRB).
\vspace{-1mm}
\subsection{Materials}
\subsubsection{Practice Tasks}

\begin{figure}
    \centering
    \includegraphics[width=0.9\linewidth]{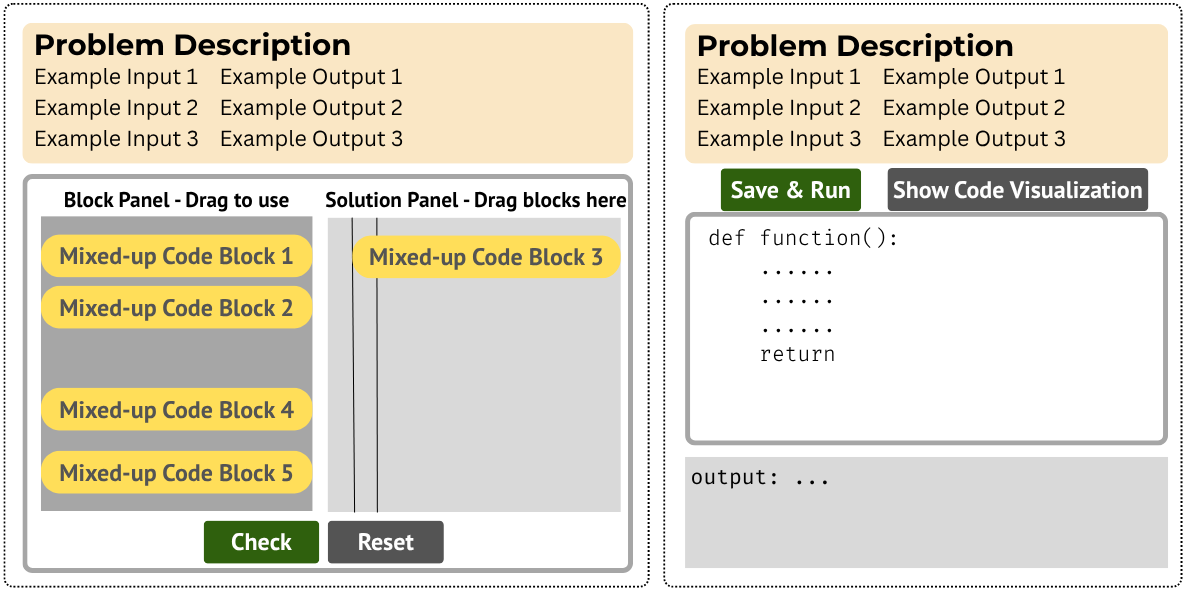}
    \caption{Two practice problem types in the study}
    \label{fig:practice}
    \vspace{-6mm}
\end{figure}

We selected two challenging programming tasks from a popular technical interview database, cross-referencing them with the course syllabus to ensure an appropriate level of difficulty. The tasks were designed to be difficult enough to encourage students to naturally use GenAI for assistance, rather than solving them independently. To investigate how programming task formats affect students' modality choices, we made one task into a mixed-up code puzzle and another into a short program writing task. A mixed-up code puzzle (Parsons puzzle) requires students to put all the mixed-up code blocks into a correct order to solve \cite{ericson2022parsons}. We chose a two-dimensional code puzzle that required students to position blocks correctly both vertically and horizontally to solve it (Figure \ref{fig:practice}-left). A short programming task requires students to write code from scratch and pass the built-in unit test cases to complete it (Figure \ref{fig:practice}-right). The two questions can be solved only using the concepts covered in the course.

\vspace{-1mm}

\subsubsection{Covered AI Input-Output Modalities}
\begin{figure}
    \centering
    \includegraphics[width=1.02\linewidth]{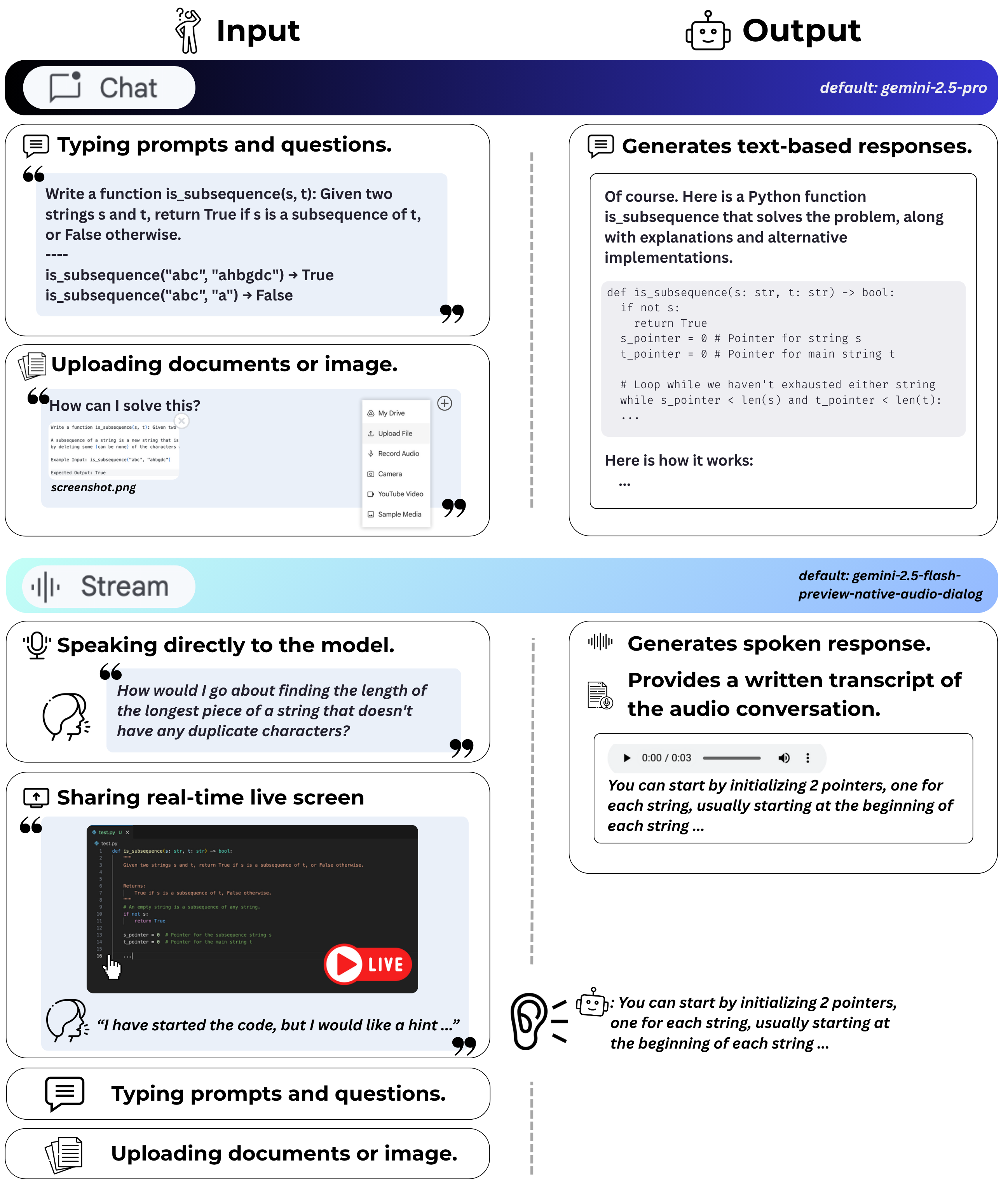}
    \caption{Main AI interaction modalities in this study. In Chat, students can use typed input or upload images and documents. AI processes input into text output. In Stream, the students can speak through a microphone, share screen, type, or upload files. AI then processes this information stream to produce a dynamic output with audio and a transcript.}
    \label{fig:modalities}
    \vspace{-12mm}
\end{figure}

We outlined diverse input-output modality choices in the context of programming problem-solving (Figure \ref{fig:modalities}). We selected Google AI Studio to carry out our study as it offered a comprehensive suite of modalities. The primary input modalities we introduced to students included: (1) \textit{text}, allowing students to type queries, including natural language and/or code; (2) \textit{file upload}, where students can upload images (e.g., screenshots, focus of this study given the question type) or documents; (3) \textit{speech}, allowing students to speak directly into a microphone; and (4) \textit{screen-sharing}, where students can manipulate on-screen content (e.g., by highlighting, scrolling, or editing) and speak at the same time. The output modalities included: (a) \textit{text}, written answers in the AI user interface, often with formatted code; and (b) \textit{audio (with written transcripts)}, which delivers an audio response with a real-time written transcript in the AI user interface.

\subsection{Procedure}
The study started after verifying eligibility and obtaining consent to record. Participating students first completed a survey on their use of generative AI. Recognizing that multimodal GenAI is still new to most students, we then included a warm-up session to help students familiarize themselves with different AI modalities. After a 5-minute introductory video, students tackled two warm-up examples that used the same format as the primary practice but covered different topics. Researchers encouraged students to engage with every modality when working with the example tasks, followed by a short think-aloud training.

Following the warm-up session, students engaged in the main practice session. They were asked to complete two programming tasks on an online learning platform (one mixed-up code puzzle and one short programming task) using AI Studio while thinking aloud. Students were allowed to choose whatever input and output modalities they preferred during this practice. Participants were given approximately 15 minutes per task but were instructed to use the AI for assistance in whatever way felt most natural, rather than focusing on completing the task. Due to time constraints, we did not have participants solve another long, open-ended programming task in this session. Instead, we asked them to bring a completed project from the course (including the project description and their code) that they found most challenging. Students were then asked to compare their past experience with the project with the potential benefits of multimodal GenAI. Then, we conducted a semi-structured interview about their practice experience, focusing on modality choices and corresponding reasons. The study ended with a short demographic survey. The study protocol was tested and refined with three pilot testings. After that, the study was conducted and recorded with a total of 16 participants.

\vspace{-2mm}

\subsection{Data Analysis}
To answer RQ1, we applied thematic analysis to analyze students' behaviors using inductive coding \cite{ando2014achieving}. We analyzed the video recordings and their transcripts. Two researchers first familiarized themselves with the data and generated an initial set of codes. They then created a codebook and independently coded the materials, resulting an IRR of cohen's kappa > 0.66 (substantial agreement \cite{mchugh2012interrater}) for the independent coding results. After that, they met to discuss and resolved all the disagreements. Results are presented in Section 4.1. For RQ2, we conducted a reflexive thematic analysis \cite{braun2019reflecting} to understand students’ criteria for choosing modalities. Two researchers went through all relevant data to develop initial themes. They then met to discuss, refine, and name each theme, they also revisited the data to identify additional supporting quotes. The results are presented in Section 4.2, with each \textbf{bolded phrase} indicating a distinct theme. For reporting, students were referenced by their PID (e.g., P1 to P16) to maintain anonymity and consistency.

\vspace{-3mm}

\section{Results}
We have 16 participants, with 6 identified as women, 9 as men, and 1 undisclosed. Students showed a diverse language background. Although English is the official language at the institution, the primary language spoken at home varied: 3 spoke only English, 4 spoke primarily English, 4 spoke another language primarily, with five bilingual. All participants had experience with commercial GenAI tools, such as ChatGPT, Gemini, Claude, or GitHub Copilot.
\vspace{-1mm}
\subsection{What \& How: Students' Input-Output Modality Choices}
While solving programming tasks with multimodal GenAI, students needed to make modality choices (intentionally or unintentionally) in three main aspects: 1) \textit{context input modality}: choose a modality to give AI their current context information; 2) \textit{instruction input modality}: choose a modality to let AI know their needs; and 3) \textit{output delivery modality}: choose a modality to obtain the AI-generated information. We first reported the context input modalities chosen by the students, followed by their associated instruction input modalities and AI-generated output delivery modalities. 

Students used AI for varied goals when solving the programming tasks. We identified, categorized, and grouped unique student-AI interaction patterns. Then we reported the counts as the number of ``cases''. In this study, students used AI to understand the problem better (11 cases), to get an idea to solve the problem (7 cases), to check their existing ideas (4 cases), to get support when implementing a solution (27 cases), to verify their current solution (13 cases), and to debug (10 cases). However, we did not observe clear differences in AI usage among diverse modality choices. Therefore, instead of comparing, we mainly described how students interacted with AI through different modalities. We used \speech{speaking} to refer to students' speaking input, \behavior{behaviors} refers to observed students' behaviors, and \keyboard{typed input} refers to students' typed input.

\vspace{-1mm}

\subsubsection{Screen-Sharing.} In over half of the context input modality selections, students chose to use screen-sharing (18 cases) to input problem-solving context, including problem and progress. Of these, in 15 cases (83\%), students selected speaking as their instruction input modality. For example, after reading the description and dragging three blocks, P2 used AI to understand the problem better:

\vspace{-1mm}

\begin{tcolorbox}[enhanced,colback=gray!8,boxrule=0pt,frame hidden, arc=4pt,top=2pt, bottom=2pt, left=6pt, right=6pt, boxsep=0pt]

    \behavior{P2 shared the screen with problem description, examples, and progress} P2 (speaking): \speech{Can you help clarify this question, and on example two?}
    
    AI (audio): ``The question is asking to find the maximum number of vowel letters within any substring of a given string s...''
\end{tcolorbox}

\vspace{-2mm}

With the shared screen, in 3 cases (17\%), students chose typing as the input modality for instruction. A notable example was P5, who switched from speech to type in the middle of problem-solving and explained it as \textit{"I may want to type so I have more time to form question."} Once they had written and tested the code, P5 used AI to get hints to continue on an unfinished solution:

\vspace{-2mm}

\begin{tcolorbox}[enhanced,colback=gray!8,boxrule=0pt,frame hidden, arc=4pt,top=2pt, bottom=2pt, left=6pt, right=6pt, boxsep=0pt]
    \behavior{P5 shared the screen with problem description, examples, and progress}  P5 (typing): \keyboard{I have started the code but I would like a hint as to how to go about the order (please ignore the "abc cdafgb abc" on the side of the screen, as that was just temporary)}. 

    AI (audio): `` You could try using two pointers'' \behavior{P5 closed the audio, but carefully read the transcript} 
\end{tcolorbox}

\vspace{-2mm}

For the output modality to receive AI answers, in 15 cases (84\%), the students chose to listen to a human-like voice audio while looking at the shared practice task screen and made modifications on that screen. For example, after finishing dragging some blocks to the solution area, P9 asked AI when they came up with a partial solution, but did not know how to continue:
\vspace{-1mm}

\begin{tcolorbox}[enhanced,colback=gray!8,boxrule=0pt,frame hidden, arc=4pt,top=2pt, bottom=2pt, left=6pt, right=6pt, boxsep=0pt]
    \behavior{P9 shared the screen with problem description, examples, and progress}
    
    \vspace{0.6mm}
    
    P9 (speaking): \speech{okay, I'm doing a problem here.} \behavior{used the mouse to highlight the problem description area} \speech{and the prompt is provided here}. \behavior{mouse circulated the right solution panel} \speech{and this is what I have so far on the right, am I on the right track, or what should I be doing?}
                
    AI (audio): ... \behavior{P9 looked at the practice task screen}
    
    P9 (speaking): \speech{Okay and then} \behavior{dragged two blocks to the solution panel} \speech{if I make this change, am I doing the right thing here? would it decrease the vowel count?}

    AI (audio): ... \behavior{P9 looked at the practice task screen, mouse idle}
\end{tcolorbox}

\vspace{-2.5mm}

At times, students chose to switch to the Stream interface and read the paired textual transcript. Some still kept the audio on, but some closed the audio and just read the transcript (as P5 above). After reading the problem description, P4 used AI to get some ideas and also a code solution. While sharing the screen on the problem description, P4 returned to the Stream interface to read the textual transcript while listening to the human-like AI audio:

\vspace{-1mm}
\begin{tcolorbox}[enhanced,colback=gray!8,boxrule=0pt,frame hidden, arc=4pt,top=2pt, bottom=2pt, left=6pt, right=6pt, boxsep=0pt]
    \behavior{P4 shared the screen of the problem description and the code example} P4 (speaking): \speech{Can you hear me?}

    AI (audio): ``Yes, I can hear you clearly. How can I help you today?''
    
    P4 (speaking): \speech{Great. So I want to write this function. And I'm wondering how I can start it.}

    AI (audio): ``You can start by initializing 2 pointers, one for each string, usually starting at ... Then you can ...''

    P4 (speaking): \speech{So can you show me how to write that in code?}

    AI (audio): ``Sure, you can initialize 2 pointers, say i and j to 0, then use ...'' \behavior{P9 looked at the AI Stream interface, and read the transcript with continuous mouse movement}
\end{tcolorbox}

\vspace{-1.5mm}

In two cases, students even copied the code lines from the transcript to their problem area. For example, P3 found a detail in the corresponding transcript and copied it to their code entry box:

\vspace{-1.5mm}

\begin{tcolorbox}[enhanced,colback=gray!8,boxrule=0pt,frame hidden, arc=4pt,top=2pt, bottom=2pt, left=6pt, right=6pt, boxsep=0pt]

    P3 (speaking): \speech{Help me finish the slicing approach then.} \behavior{used the mouse to highlight two code lines} \speech{does that make sense or should I change it to something else?}

    AI (audio): ``... How about we use the find method ...?''
    
    P3 (speaking): \behavior{used the mouse highlighted one code segment char in string s} \speech{Ok, what should I change this to then?}

    AI (audio): ``You can change the if char ... For example, you could try if char ... \behavior{P3 started to type to follow AI's suggestion as\hspace{0.3mm} \keyboard{char.find(char)???} in the practice task} does not equal negative one, this will check ...''

    \behavior{P3 switched back to AI Stream interface and went through the transcript using the mouse. P3 copied \keyboard{``if s.find(char)!=-1''} in the transcript to replace their \keyboard{``if char.find(char):''} line in the practice task}

    P3 (speaking): \speech{how does the find method work?}

    AI (audio): ``The find method returns the index of ...''
\end{tcolorbox}

\vspace{-2mm}

We observed instances where students conducted multiple-round voice conversations with AI, while sharing a screen as the contextual foundation. Students could ask clarification questions about details in the AI reply (like P3 above), release the next request (like P4 above), and even check their new progress after following AI's suggestions. For example, P8 checked their new modification after listening to AI's suggestion:

\vspace{-2mm}

\begin{tcolorbox}[enhanced,colback=gray!8,boxrule=0pt,frame hidden, arc=4pt,top=2pt, bottom=2pt, left=6pt, right=6pt, boxsep=0pt]
    AI (audio): ``You can start by using 2 pointers, one for each string ...'' \behavior{P8 wrote some code in the practice task screen}

    P8 (speaking): \behavior{looked at the practice task screen} \speech{like this?}

    AI (audio): ``Yes, that's a good start. You'll need to add the logic to compare the characters and move the pointers accordingly''

    -- --
    
    P8 (speaking): \behavior{looked at the practice task screen} \speech{What do you mean by pointer?}

    AI (audio): ``I meant 2 pointers to keep track of the current position you're looking at in each string ...'' \behavior{P8 wrote two lines in the practice task screen}

    P8 (speaking): \behavior{looked at the practice task screen} \speech{like this?}

    AI (audio): ``Exactly. Now, inside the loops, you'll need to  ...''

\end{tcolorbox}

\vspace{-3mm}

\subsubsection{Screenshot Image.} In 10 cases (around 30\%), students uploaded screenshot images to capture their problem-solving context (e.g., problem description, code progress, and the error feedback). As for instruction input modality, only one student (P6) provided a screenshot of the problem description. Others combined screenshots with typed instruction prompts to guide the AI, but might not do so in a single input attempt. For example, after reading the description, P1 first sent a problem description screenshot image, quickly skimmed the AI response, and then typed a follow-up question as \keyboard{"how would you proceed to code this?"}. After completing a solution, P1 took another screenshot of the current solution, and typed as \keyboard{"is this a correct implementation?"}.

Two output delivery modalities were available for this input: the audio with textual transcript and pure textual output. Observed in 5 of 12 screenshot-input outputs (42\%), the students chose to listen to audio with transcripts. For example, P10 screenshoted their current block arrangement and received AI audio. After listening to the audio, P10 read the transcript again and backed to move two block. Then P10 sent a new screenshot with the updated block arrangement to check its correctness. In the rest output cases (58\%), the students chose to read the pure textual output. 

\vspace{-2mm}

\subsubsection{Type or Paste.}  With other modalities available, not many use the traditional modality of typing or copy-pasting context. However, in around 14\% of the times (5 cases), students either pasted context into the input field or skipped this step. In such cases, the students always typed their instructions. For instance, after putting some blocks in the solution panel, P14 copied the problem description, pasted and typed as \keyboard{"[Problem description], Explain exactly what is being asked for me to do without generating the actual code> The example inputs are: [the given examples]"} We only observed one case where the student (P12) typed their instructions but chose to receive the AI-generated output via audio, without clear sign of reading. In other times, students read and even copy the textual AI-generated output.

\vspace{-3mm}

\subsection{Why: Factors Influencing Modality Choices}
We structured the results by first presenting shared factors that influenced both input and output modality decisions, followed by reasons specific to input modality choices, and then the reasons associated with output modality choices.

\vspace{-2mm}

\subsubsection{Universal Considerations in Selecting Input and Output Modalities}

\textbf{The perceived time to reach a goal.} One main reason that influences students' overall modality choice is the perceived time investment for a given AI modality combination for a given goal. For example, P15 found time pressure to be one main factor for using screenshot, \textit{"when I take that screenshot, I can let AI do its own thing while I could possibly be working on something else. That's definitely a big time saver, especially if I'm working on something that has a very close deadline."} P11 preferred speech over type because \textit{"it helps me communicate faster, rather than type out my answers"}. 

\textbf{The physical setting and devices.} This is a key factor in determining whether students opt for auditory-related modalities. For instance, P2 emphasized transcript as an important complement to audio output because \textit{"there might be an issue with my audio, maybe my headphones, or my earphones, or maybe my laptop speakers, and I think that this transcript kind of supplements that and basically eliminates any of those issues"} P7 treated audio output as a complement to textual output as when reading was impractical, \textit{"if I don't have the time to read the text, and then it (audio output) would be helpful, like when I'm on the go, I'm on the bus and I want to hear it."} 

\textbf{The degree to which AI simulates human.} Screen-sharing, speech input with audio output was favored by students who valued AI's human-like tutoring: \textit{"I'm kind of connecting this to real life what I would do with like my friends or peers, or maybe instructors as well."} (P15) And P7 favored the speech input because it \textit{"almost like interacting with a human being."} Nevertheless, this does not mean AI interactions should fully mimic real human conversation, since students still wanted more dominant control. As P2 noted, \textit{"I wish it's like push to talk instead of constantly on."} P16 was desatisfied with the inability of interrupt an AI output, indicated by a series of "no"s (\textit{"Ouch, no, no, no, no, no. Hold on."}), clearly signaling dissatisfaction.

\textbf{Personal preference for learning.} Sometimes students just chose the modality they personally favored. For example, P13 preferred typing over speaking as input because \textit{"I have a little bit of a paranoia with random sites who are taking voice input."} Conversely, P15 preferred audio than textual output considering \textit{"me personally, I don't really like looking at really long articles like Wikipedia articles. I find that really hard to digest. I like to listen to information and kind of soak it up through that way."}

\vspace{-2.5mm}

\subsubsection{Reasons Specific for Input Modality}
\textbf{Seamless versus intentional spacing.} Preferred intervals between AI rounds varied. Students who valued seamless interaction liked the use of screen-sharing and microphone to speak, \textit{"...talking to it is much better than typing out, because maybe you're thinking of something, so you just want to try to say it out to it and then it'll capture what you're trying to say at that moment in time."} (P2) Conversely, some preferred typing input for the time afforded to formulate their questions. As P6 raised, \textit{"typing out my problems, gives me the opportunity to really go into detail and think about it more"}. P6 also reported a feeling of rush caused by speaking and screen-sharing, \textit{"it has to be a bit more rushed, and I have to come up with things more on the spot. So I prefer text and being able to upload screenshots."} 

\textbf{Possible redundant effort for students.} Screen-sharing with speech input allows the continuous context updates, which was valued by students who wanted to minimize effort to provide repeatedly changing information. As P2 described, \textit{"I don't have to switch screens. I think with LLMs, you have to switch and then type out your entire message before you can get an answer."} P12 even flagged the potential to lose context when using traditional typing input, \textit{"before they have in this feature, I have to like copy, paste and might even lose some data in this process."}

\textbf{Scope of presentation format coverage.} Screen-sharing and screenshot images afforded more information formats than typing input. P12 pointed it out as \textit{"if I have text in the image that I can't copy, I could just take a screenshot and send the whole thing."} P9 also valued screen-sharing's broader format coverage scope for mixed-up code puzzles, \textit{"this was about placing blocks, and this wasn't just text I could copy directly. It was more about the orientation position of my block, so something like ... visual thing ... it makes a lot more sense if the AI could see my screen in real time."} 

\textbf{Difficulty with natural language articulation.} When a clear textual description was challenging, students favored other modalities over pure text input. According to P14, \textit{"A photo is useful in cases when you need help on concepts you do not understand well enough to explain for proper assistance"} 

\vspace{-1.5mm}

\subsubsection{Reasons Specific for Output Modality}
\textbf{Standards of a good AI answer.} What students prioritized in AI answers influenced their output choice. For example, P3 liked audio output for ideation, \textit{"for ideas, the audio was like the best. Because when you're thinking ... it kind of keeps up with your thought process."} But when it comes to syntax that requires higher precision, many wanted textual output,  \textit{"for that (syntax) I found the text helpful. It would have been even more helpful if it was able to generate like code blocks."} (P3)

\textbf{Opportunity to do asynchronous revisit.} Compared to audio output, textual output allowed for rereading and closer digest, \textit{"put out the voice as well as transcript, even though I could hear the answer, I would like go back and read it, because that's sort of what sticks with me more, and I can just go back and check, highlight certain things, or copy-paste."} (P6) and \textit{"if I forgot what it said or I wanted to go back and review, the transcript was just there, right below the recording (audio), so that was helpful."} (P4) 

\vspace{-2mm}
\section{Discussion and Future Work}
Moving from text-focused to a more flexible multimodal interaction with GenAI, this work enriched the modality dimension of using GenAI in programming learning. As one of the early efforts in this area, this study reveals students’ initial perceptions among various modality options. It opened up new directions for future work.
 
We found that students have varied preferences for modalities when using AI to tackle programming problems. Some particularly appreciated the combination of screen sharing, voice input, and audio output, as it closely resembled the interactive dynamic of human tutoring, much like virtual office hours \cite{malan2009virtualizing,gao2022uses}. This aligns with earlier efforts to use text-to-speech for human-like voice AI tutors in multimedia learning \cite{ liew2023alexa,craig2017reconsidering}. They found that a human-like voice effectively facilitates social connections with learners and decreased cognitive load \cite{liew2023alexa}. However, some students still favored more spaced interaction with AI (screenshot images and typing as input) and textual output. One obvious reason is that many commercial AI tools still focus on text modalities. Our study may be their first time using other modalities, so they were not yet familiar with them. Furthermore, students were self-regulated \cite{zimmerman2011self} in knowing which modality is the most appropriate under different learning situations. Some valued the intentional spacing brought by typing input to formulate their thinking, and some strategically selected output modalities depending on whether they were at ideation or detail processing stages. Their personal learning preference also impacted the modality choices \cite{truong2016integrating}. 

For multimodal complementarity, even when students chose to use the synchronous modalities, many emphasized the importance of the textual transcript for asynchronous review. This is because transcripts have been shown to aid understanding during meetings \cite{gao2014effects} and improve instructional video comprehension \cite{grgurovic2007help}. 

This work serves as a starting point to \textbf{open new directions to integrate multimodal GenAI in CS education}. From an educator's perspective, as students hold different modality preferences, it is worthwhile for instructors to introduce different GenAI modalities to students to fit their learning preferences. In addition, instructors should choose the modalities carefully when integrating GenAI into instructional activities. From a tool design perspective, it is important to give students the agency to shift among different modality combinations. Moreover, some students found that sometimes AI talks too much, or students could not disturb or correct their meaning. We recommend giving students more control over the conversation flow, such as the ability to initiate the dialogue on their own terms and to interrupt it mid-response.

Furthermore, while we focus on the use of multi-modality GenAI in programming learning, we envision its extension beyond programming learning. Auditory-based modalities are suitable for technical communication training, especially for conversational programmers \cite{chilana2016understanding}, while multi-modalities can provide a foundation for diverse real-world training when AI takes different roles; (1) technical interview preparation when AI is asked to assume roles such as interviewers; and (2) CS instructor training and TA training to manage office hours \cite{ball2021teaching} where AI takes the role of students. 

\vspace{-5mm}

\section{Limitations}
This study involved a small group of undergraduate students from a single institution, which limits its generalizability across contexts. Also, we focused on short Python programming questions; future work should examine more complex, open-ended tasks and other programming languages. In addition, as participants were new to multiple GenAI modalities, their behaviors may change with routine use. Finally, we lacked sufficient evidence on students’ personalization of the AI’s voice or language, which may be especially relevant for non-native English speakers.
\vspace{-4mm}

\section{Conclusion}
This work explored how students used multimodal GenAI to solve programming problems. We observed varied preferences on students' modality choices with diverse underlying criteria. Some preferred AI interaction that felt more seamless and human tutoring, while others remained with traditional text-based interaction. The underlying reasons included the nature of the task (e.g., time given, physical environment, good AI answer standards) and students' personal preferences (e.g., seamless communication vs. intentional pause and spacing for articulating questions). These insights show key directions for future learning experience design: enabling learners to seamlessly switch modalities, informing AI-driven educational tools across diverse contexts, and predicting a new era of multimodal AI-enhanced education.

\bibliographystyle{ACM-Reference-Format}
\bibliography{references}

\end{document}